# Superstatistics of Schrödinger Equation with Pseudoharmonic potential in an External Magnetic and Aharanov-Bohm(AB) Fields.

A.N.Ikot [1], U.S.Okorie[1&2], G.Osobonye[3], P.O.Amadi[1],
C.O.Edet*[1], G.J.Rampho[4] and R.Sever[5]

[1]Department of Physics, University of Port Harcourt, Port Harcourt- Nigeria.
[2]Department of Physics, Akwa Ibom State University, Ikot Akpaden, Uyo.-Nigeria
[3]Department of Physics, Federal College of Education, Omuku, Rivers State- Nigeria
[4]Department of Physics, University of South Africa.
[5]Department of Physics, Middle East Technical University, 06800, Ankara, Turkey.

**Abstract**
In this work, the thermodynamic property of pseudoharmonic potential in the presence of external magnetic and AB fields is investigated. We used effective Boltzmann factor within the superstatistics formalism to obtain the thermodynamic properties such as Helmholtz free energy (F), Internal energy (U), entropy(S) and specific heat (C) of the system. In addition, we discuss the result of the thermodynamic properties of some selected diatomic molecules of $N_2, Cl_2, I_2$ and $CH$ using their experimental spectroscopic parameters and that of the variation of the deformation parameter of $q = 0, 0.3, 0.7$. We also illustrated with some graphs for clarity of our results in both cases.

*Corresponding author: collinsokonedet@gmail.com

# 1. Introduction

The solutions of the Schrödinger equation and its relativistic counterpart with different physical potential models plays an imperative role in many fields of physics and quantum chemistry since these solutions contain all the necessary information required to describe a quantum system[1-15]. Furthermore, the solution of the Schrödinger equation is of great importance in particle, nuclear and chemical physics amongst other. Its solution can be used to investigate the mass spectra, binding energies, decay rates transition properties and thermodynamic functions [16-18].

In quantum mechanics only few physical systems can be solved exactly in a closed form, for example the harmonic oscillator and Coulomb potential [19-20], pseudoharmonic and Kratzer potential[21-23] etc. The solutions of the Schrödinger equation in a 2D charged particles confined by a harmonic oscillator in the presence of external magnetic field along the z-axis and Aharonov Bohm(AB) flux field created by solenoid have been studied by many authors[24-27]. Also, other effects of external magnetic fields on different physical systems have been investigated [28]. Koscik and Okopinska investigated the quasi exact solutions for two interacting electrons using Coulombic force with confined anisotropic harmonic oscillator in 2D anisotropic dot[29]. Aygun et al.[30] studied the effect of constant magnetic field on the energy spectra of a particle moving under the Kratzer potential using asymptotic iteration method(AIM)[31]. Also, Ikhdair and Hamzavi[32] studied the spectral properties of quantum dots via Schrodinger equation with anharmonic potential and superposition of pseuodoharmonic-linear-Coulomb potential in the presence of an external uniform magnetic and AB flux fields using Nikiforov-Uvarov(NU) method[33]. Furthermore, the solutions of the Dirac equation with anharmonic potential in the presence of external and AB fields have been obtained [34]. Subsequently, many Schrödinger-like equations with harmonic and annharmonic potentials with and without external magnetic and AB flux fields has been investigated [35-37].

In addition, the determinations of the thermodynamic functions of gases and diatomic molecules over a wide range of temperature limits have attracted the attention of many researchers [38]. Yepes et al.[39] studied the heat capacity and magnetization for GaAS quantum dot with asymmetric confinement. The thermal and magnetic properties due to electronic confinement have attracted significant interest such as electron-electron interaction on the energy spectrum [40] and electronic structure [41]. For instance, Atoyan et al.[42] studied relativistic spinless particles in a 2D cylindrical potential. Gumber et al.[43] studied a 3D cylindrical QD in the presence of external electric and magnetic field and went further to determine canonical partition function and other thermodynamic properties. It is well known that once the partition function of a system is calculated then other thermodynamics properties of the system can be evaluated. The partition function which is a function of temperature is usually regarded as the distribution function in statistical mechanics which was first initiated by Boltzmann in 1870[44]. Other statistical mechanical representations besides Boltmann had been proposed such as Gibbs[45],Einstein[46],Boltzmann-Gibbs(BG)[47],Tsallis[48] and the latest which is the superstatistics[49-50].It was Wilk and Wlodarczyk [49] that first conceived idea of superstatistics before Beck and Cohen[50] latter reformulated the theory. The superstatistics described a non-equilibrium system with stationary state and intensive parameter fluctuations. The superstatistics has many applications in different branches of physics chemistry such as cosmic rays [51], wind velocity fluctuations [52], hydrodynamic turbulence [53] among others.

The aim of the present work is to investigate the Schrödinger equation in 2D for a pseudoharmonic oscillator in the presence of external and AB fields with applications to superstatistics. Pseudoharmonic potential is one of the exactly solvable potential in physics [54]. The pseudoharmonic potential is defined as,

$$V(r) = Ar^2 + \frac{B}{r^2} + C \quad (1)$$

where $A, B$ and $C$ are potential constants. As an application, we present the thermal properties of the system within the superstatistics formalism. The thermodynamic properties of some selected diatomic molecules of $N_2, Cl_2, I_2$ and $CH$ [55] within framework of superstatistics are also reported.

## 2 Solutions of the Schrödinger equation in external magnetic and AB fields

The Schrödinger equation for a charge particles moving in a magnetic and AB flux fields is[24-27],

$$\left[\frac{1}{2\mu}\left(\vec{P} + \frac{e}{c}\vec{A}\right)^2 + V(r)\right]\psi(r,\varphi) = E\psi(r,\varphi) \quad (2)$$

where the vector potential $\vec{A}$ can be expressed as $\vec{A} = \vec{A}_1 + \vec{A}_2$ in such a way that $\vec{\nabla} \times \vec{A}_1 = \vec{B}$ and $\vec{\nabla} \times \vec{A}_2 = 0$ with $\vec{B} = B\hat{z}$ being the applied magnetic field and $\vec{A}_2$ describing the additional magnetic flux $\Phi_{AB}$ created by a solenoid[24-27],

$$\vec{A_1} = \frac{1}{2}(\vec{B} \times \vec{r}) = \frac{Br}{2}\hat{\phi}, \vec{A_2} = \frac{\Phi_{AB}}{2\pi r}\hat{\phi},$$

$$\vec{A} = \vec{A_1} + \vec{A_2} = \left(\frac{Br}{2} + \frac{\Phi_{AB}}{2\pi r}\right)\hat{\phi} \quad (3)$$

where $\Phi_{AB} = \frac{2\pi}{e}$.

Now substituting equation (1) into equation (2), we get

$$\left[\nabla^2 - \frac{e^2}{\hbar^2 c^2}A^2 - \frac{e}{\hbar^2 c}\vec{P}.\vec{A} - \frac{e}{\hbar^2 c}\vec{A}.\vec{P}\right]\psi(r,\varphi) + \frac{2\mu}{\hbar^2}\left[E - Ar^2 - \frac{B}{r^2} - C\right]\psi(r,\varphi) = 0 \quad (4)$$

Now inserting equation (3) into equation (4), we obtain

$$\left[\nabla^2 - \frac{e^2}{\hbar^2 c^2}\left(\frac{Br}{2} + \frac{\Phi_{AB}}{2\pi r}\right)^2 - \frac{eB}{2\hbar^2 c}\frac{\partial}{\partial\varphi} - \frac{e\Phi_{AB}}{2\pi\hbar^2 c}\frac{1}{r^2}\frac{\partial}{\partial\varphi}\right]\psi(r,\varphi)$$
$$+ \frac{2\mu}{\hbar^2}\left[E - Ar^2 - \frac{B}{r^2} - C\right]\psi(r,\varphi) = 0 \quad (5)$$

In order to find an exact solution of equation (5), we make the following choice for the wave function as,

$$\psi(r,\varphi) = \frac{R(r)}{\sqrt{r}}e^{im\varphi}, m = 0, \pm 1, \pm 2... \quad (6)$$

and this transform equation (5) into the second order Schrödinger equation as

$$\frac{d^2 R(r)}{dr^2} + \left\{\frac{2\mu E}{\hbar^2} - \frac{2\mu}{\hbar^2}\left(Ar^2 + \frac{B}{r^2} + C\right) - \frac{1}{r^2}\left(m^2 - \frac{1}{4} + \frac{e^2\Phi_{AB}^2}{4\pi\hbar^2 c^2} - \frac{em\Phi_{AB}}{2\pi\hbar c}\right) + \frac{e^2 B^2}{4\hbar^2 c^2}r^2\right\}R(r) = 0. \quad (7)$$

We can further expressed equation (7) as,

$$\frac{d^2 R(r)}{dr^2} + \left\{-\alpha r^2 + \frac{\gamma}{r^2} - \varepsilon^2\right\}R(r) = 0 \quad (8)$$

Where,

$$\varepsilon^2 = -\frac{2\mu}{\hbar^2}(E - C),$$

$$\alpha = \left(\frac{2\mu A}{\hbar^2} - \frac{e^2 B^2}{4\hbar^2 c^2}\right), \quad (9)$$

$$\gamma = \left(\frac{1}{4} - m^2 + \frac{em\Phi_{AB}}{2\pi\hbar c} - \frac{e^2\Phi_{AB}^2}{4\pi^2\hbar^2 c^2} - \frac{2\mu B}{\hbar^2}\right)$$

To get the solution of equation (8), we introduce a coordinate transformation of the form $s = r^2$ and equation (8) becomes,

$$\frac{d^2 R(s)}{ds^2} + \frac{1}{2s}\frac{dR(s)}{ds} + \frac{1}{s^2}\left\{-\frac{\alpha}{4}s^2 - \frac{\varepsilon^2}{4}s + \frac{\gamma}{4}\right\}R(s) = 0 \quad (10)$$

Equation (10) can be solve by Nikiforov-Uvarov method[56] (see appendix A) and the energy spectrum is obtained as,

$$E = \frac{\hbar^2}{\mu}\sqrt{\frac{2\mu A}{\hbar^2} - \frac{e^2 B^2}{4\hbar^2 c^2}}\left(2n+1+2\sqrt{\frac{1}{16}+\frac{\gamma}{4}}\right) + C \quad (11)$$

## 3 Boltzmann factor in the Superstatistics

The q-deformed superstatistics is a superposition of multiple different statistical models [11-15] and has many applications in physics. The effective Boltzmann factor is obtained as[11-15]

$$B(E) = e^{-\beta E}\left(1 + \frac{q}{2}\beta^2 E^2\right) \quad (12)$$

where $q$ is the deformation parameter and lies between, $0 \leq Q \leq 1$, $\beta = \frac{1}{k_B T}$, $T$ is the temperature, $k_B$ is the Boltzmann constant and $E$ is the energy of the quantum state. Consequently, one obtain the ordinary statistical mechanics i when $q = 0$ [15]. It is worthy to note that the thermodynamics properties of the system within the superstatistics framework valid for all values of $q$ and depends on the energy of the system.

## 4 Thermodynamic Properties with Pseuodoharmonic potential in the presence of an external magnetic field

The partition function in the superstatistics is defined as[11-15],

$$Z = \int_0^\infty B(E)\,dn \quad (13)$$

Substituting equations (11) and (12) into equation (13), we obtain an expression for the partition function within the superstatististics regime as,

$$Z(\beta) = \frac{1}{4P\beta}\left\{2e^{-\beta(2PQ+C+P)}q + 2\beta P e^{-\beta(2PQ+C+P)}q + C^2\beta^2 e^{-\beta(2PQ+C+P)}q\right\}$$

$$+ \frac{1}{4P\beta}\left\{2C\beta e^{-\beta(2PQ+C+P)}q + P^2\beta^2 e^{-\beta(2PQ+C+P)}q + 2e^{-\beta(2PQ+C+P)}\right\}$$

$$+ \frac{1}{4P\beta}\left\{4P^2Q^2\beta^2 e^{-\beta(2PQ+C+P)}q + 4P^2Q\beta^2 e^{-\beta(2PQ+C+P)}q\right\}$$

$$+ \frac{1}{4P\beta}\left\{2CP\beta^2 e^{-\beta(2PQ+C+P)}q\right\}$$

$$+ \frac{1}{4P\beta}\left\{4PQ\beta e^{-\beta(2PQ+C+P)}q + 4CPQ\beta^2 e^{-\beta(2PQ+C+P)}\right\} \qquad (14)$$

where,

$$P = \frac{\hbar^2}{\mu}\sqrt{\frac{2\mu A}{\hbar^2} + \frac{e^2 B^2}{4\hbar^2 c^2}} \qquad (15)$$

$$Q = \sqrt{\frac{1}{16} + \frac{\gamma}{4}} \qquad (16)$$

Other thermodynamic functions like free energy $F$, mean energy $U$, entropy $S$ and specific heat capacity $C$ can be obtain from the partition function of equation (14) using the following relations

$$F = -\frac{1}{\beta}\ln Z;$$

$$U = -\frac{\partial \ln Z}{\partial \beta};$$

$$S = k\ln Z - k\beta\frac{\partial \ln Z}{\partial \beta}; \qquad (17)$$

$$C = k\beta^2 \frac{\partial^2 \ln Z}{\partial \beta^2}$$

## 4 Applications

In this section, we evaluate the thermodynamic properties for different diatomic molecules of $N_2, Cl_2, I_2$ and $CH$ using their experimental values with a deformation parameter of $q=0.5$ in the presence and absence of the external magnetic fields. The second application is carried out by evaluating the thermodynamic properties for different deformation parameter of $q=0,0.3,0.7$ and $1$ in the present of the external magnetic fields.

### 4.1 Applications to diatomic molecules

Here, we consider the pseudoharmonic oscillator of the form [55],

$$V(r) = D_e \left( \frac{r}{r_e} - \frac{r_e}{r} \right)^2 \quad (18)$$

On comparing equation (1) and (18) gives the values of A,B and C as $A = D_e r_e^{-2}, B = D_e r_e^2$ and $C = -2D_e$ [55]. We also take $B = 3T, m = 1, \Phi_{AB} = 1$ and $\hbar = c = e = 1$ in the presence of the magnetic fields and zero otherwise. The experimental values of some selected diatomic molecules of $N_2, Cl_2, I_2$ and $CH$ are taken from Ref.[55] as shown in Table 1.

**Table 1. Spectroscopic Parameters of the selected Diatomic Molecules used [55].**

| Molecule | $r_e(\text{Å})$ | $\mu(amu)$ | $d_e(cm^{-1})$ |
|---|---|---|---|
| $N_2$ | 1.0940 | 7.00335 | 98288.03528 |
| $Cl_2$ | 1.9872 | 17.4844 | 20276.440 |
| $I_2$ | 2.6664 | 63.452235 | 12547.300 |
| $CH$ | 1.1198 | 0.929931 | 31838.08149 |

Now using these experimental data as our input, we plot the partition function for $N_2, Cl_2, I_2$ and $CH$ in the presence and absence of the magnetic field B in Figs.1-2 within the framework of superstatistics at $q=0.5$. As shown in Figs. 1 and 2, the partition functions in the presence and absence of the magnetic fields decreased monotonically as the inverse temperature $\beta$ is increased. The plots of the Helmoltz free energy for both cases are also shown in Figs.3-4. Fig.3 and 4 show the behaviour of the Helmholtz free energy versus $\beta$. It is observed in Fig.3 and 4 that the Helmholtz free energy increases as parameter $\beta$ is increased. Figs.5 and 6 show the internal energy of the diatomic molecules within the superstatistic formulation .As shown in Figs.5 and 6,the internal energy U of the diatomic molecules decreased with increasing $\beta$ in the presence of the external magnetic fields B and $\Phi$. The variations of the entropy with inverse temperature $\beta$ are shown in Figs.7 and 8 for the three diatomic molecules. Here, the entropy decreases with increasing $\beta$ in the presence and absence of the magnetic field. In Figs. 9 and 10, the variations of the heat capacity (C) as a function of inverse temperature are shown. Fig.9 shows that the specific heat peaked at unity and decreased with increasing $\beta$ for different diatomic molecules in the presence of the external magnetic field. However, in the absence of the magnetic similar trend is observed but this occurs at a very high temperature as compared to that of the present of the magnetic field which occurs at a lower temperature as shown in figs 9 and 10.

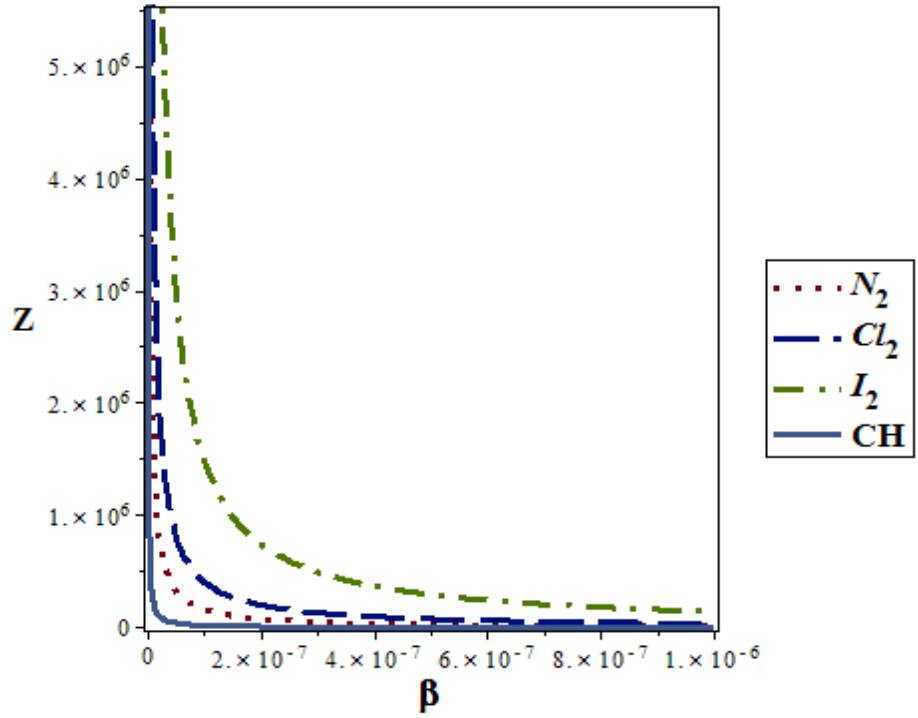

Fig. 1: Partition function vs $\beta$ for various diatomic molecules in the presence of the magnetic field B and $\Phi$

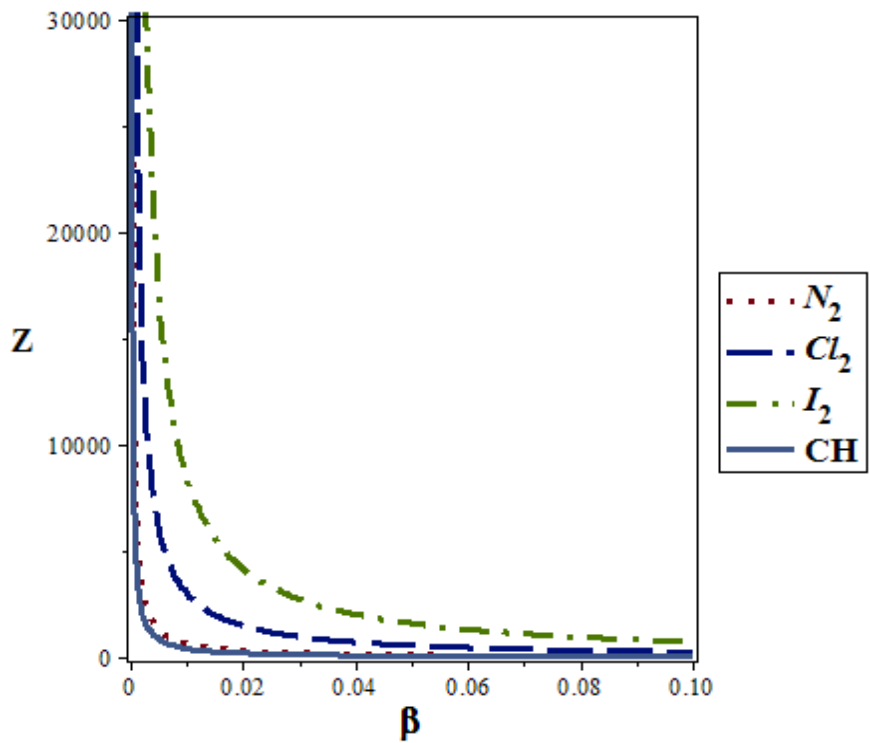

Fig. 2: Partition function vs $\beta$ for various diatomic molecules in the absence of the magnetic fields.

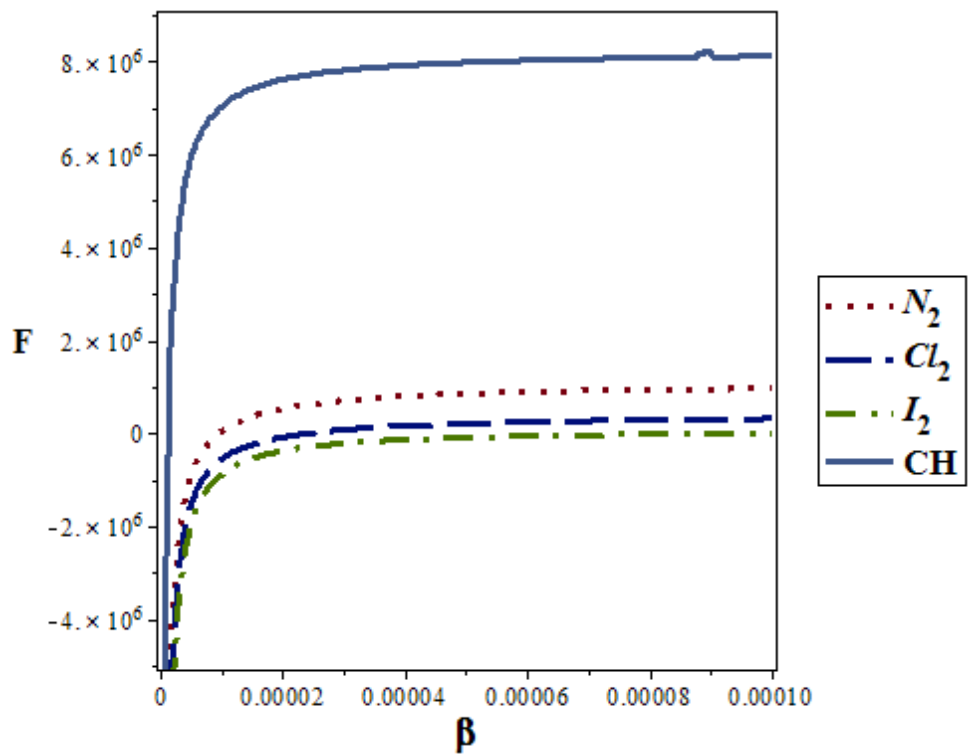

Fig. 3: Free energy vs $\beta$ for various diatomic molecules in the presence of magnetic field B and $\Phi$.

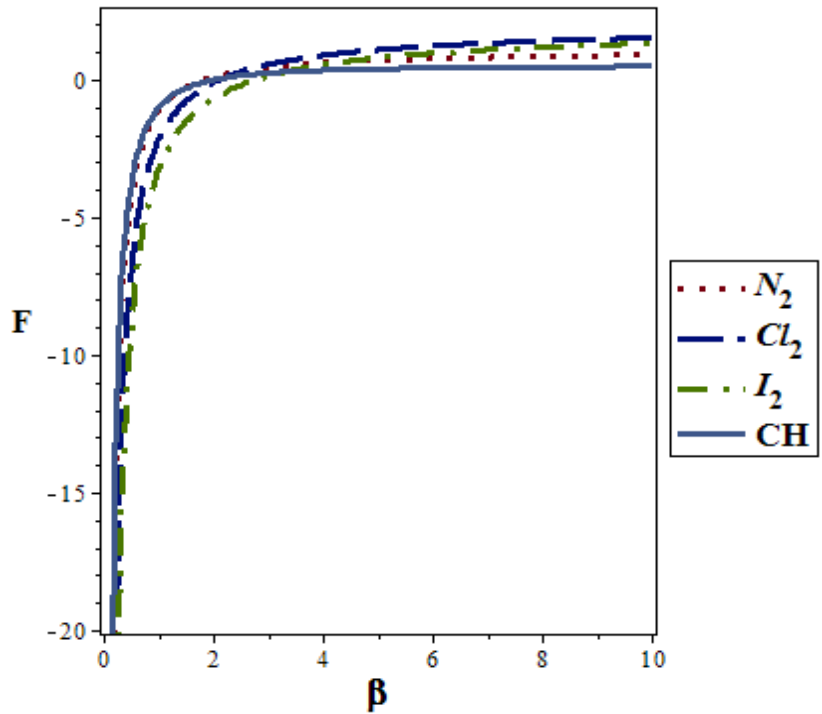

Fig. 4: Free energy vs $\beta$ for various diatomic molecules in the absence of the magnetic field B and $\Phi$.

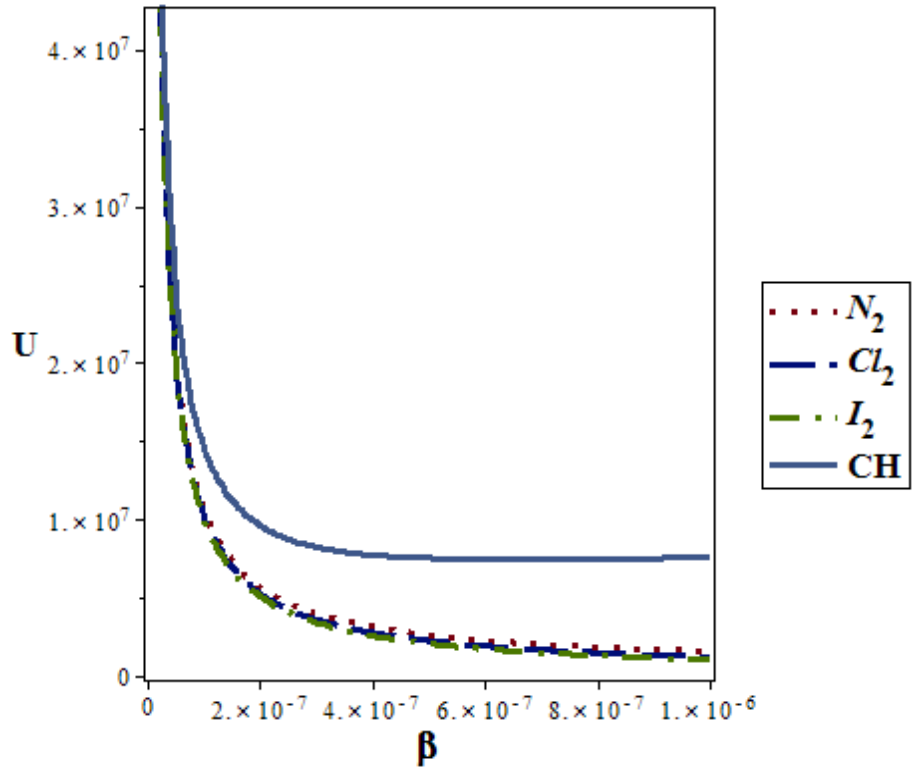

Fig. 5 : Mean energy vs $\beta$ for various diatomic molecules in the presence of magnetic fields B and $\Phi$.

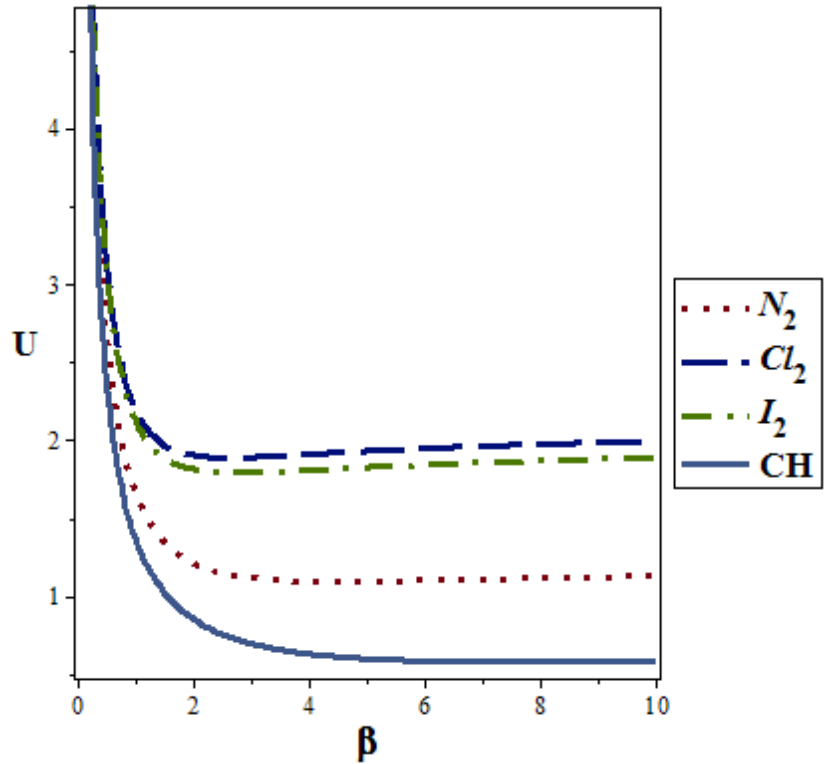

Fig. 6 : Mean energy vs $\beta$ for various diatomic molecules in the absence of the magnetic fields B and $\Phi$

.

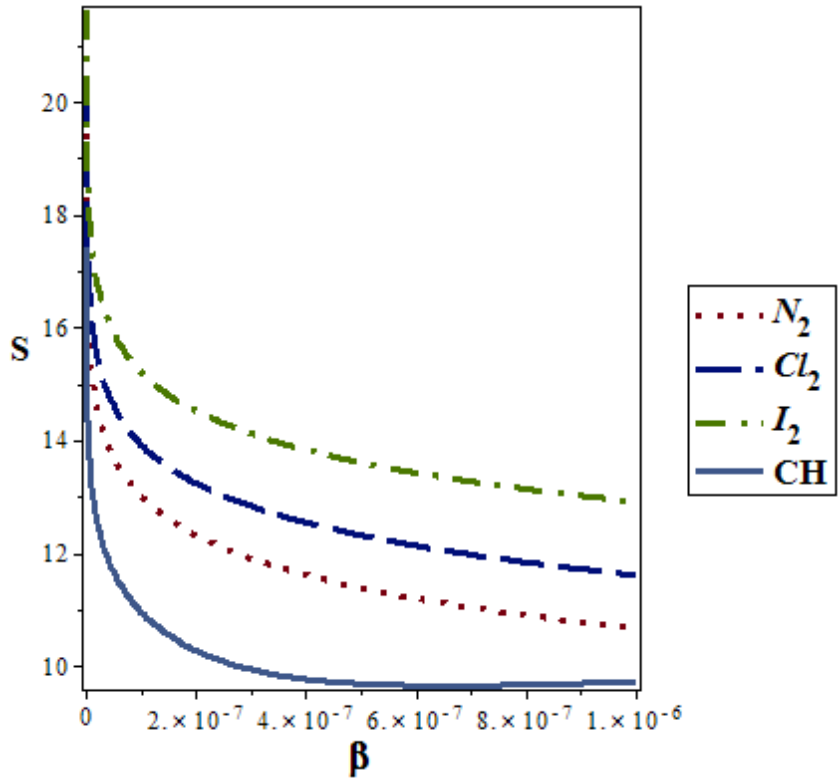

Fig. 7: Entropy vs $\beta$ for various diatomic molecules in the presence of the magnetic fields B and $\Phi$

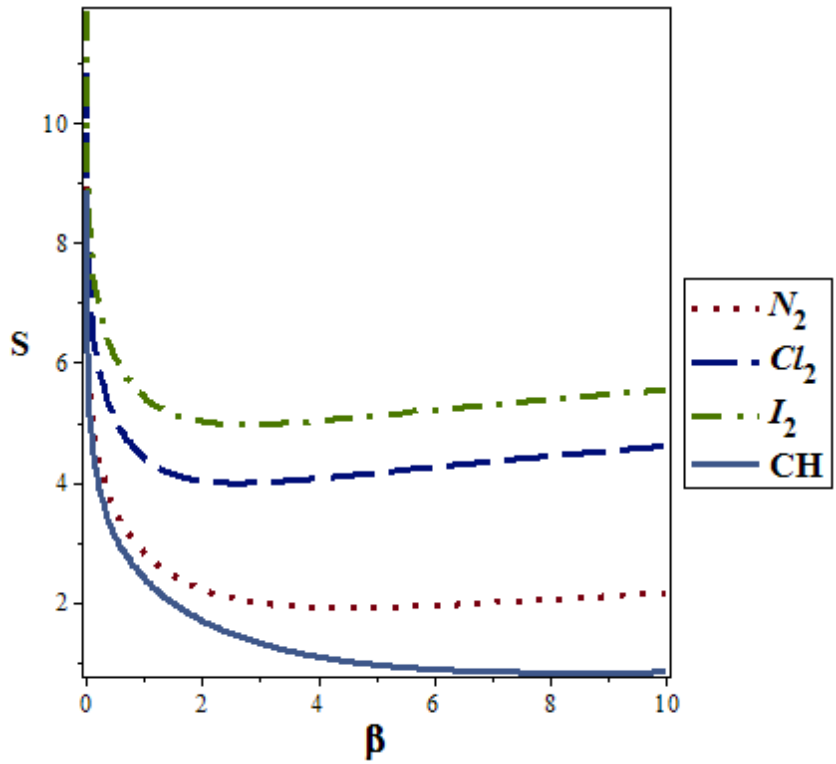

Fig. 8 : Entropy vs $\beta$ for various diatomic molecules in the absence of the magnetic field B and $\Phi$

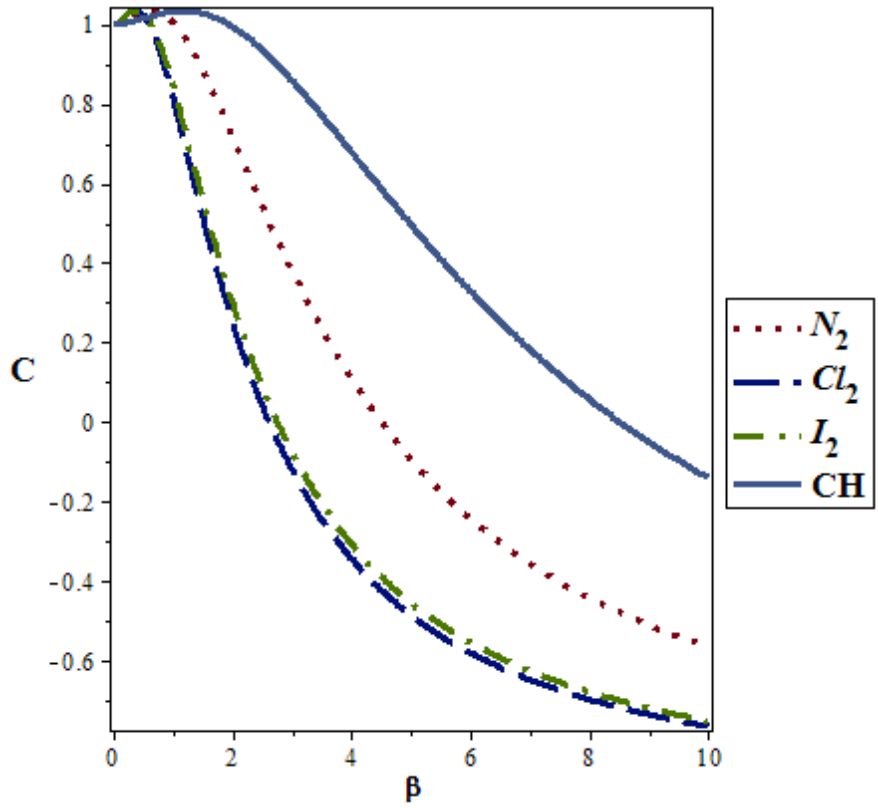

Fig. 9 : Specific heat capacity vs $\beta$ for various diatomic molecules in the presence of the magnetic field B and $\Phi$

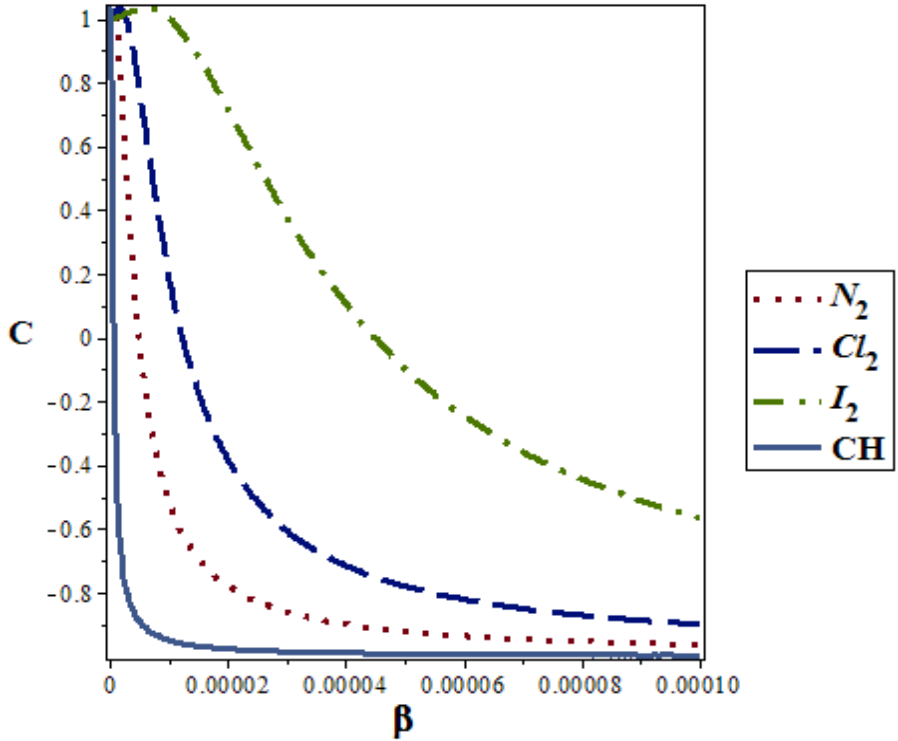

Fig. 10 : Specific heat capacity vs $\beta$ for various diatomic molecules in the absence of the magnetic field B and $\Phi$

### 4.2 Deformation parameter

Similarly, taking $B=3T, m=1, \Phi_{AB}=1$ and $\hbar=c=e=1$, we investigated the beahviour of the partition function $Z$, free energy $F$, mean energy $U$, entropy $S$, and specific heat capacity $C$ for different superstatistics parameters $q=0, 0.3, 0.7$ and 1. In fig.11, we plot behaviour of the partition function versus $\beta$ for different values of the deformation parameter $q$. As can be seen in fig.11, increasing $\beta$ decreasing the partition function monotonically. Also, in fig.12 the behaviour of the Helmholtz free energy is plotted as a function of the inverse temperature $\beta$. The Helmholtz free energy increases with increasing $\beta$ to a maximum values for different values of $q=0, 0.3, 0.7$ and 1. Fig.13 shows the behaviour of internal energy as a function of $\beta$ within the superstatistic formulation for different $q$ values. .As is observed in Figs.13, the internal energy U decreased with increasing $\beta$ in the presence of the external magnetic fields B and $\Phi$. The variations of the entropy with inverse temperature $\beta$ are shown in Figs.14. The entropy decreases with increasing $\beta$ in the presence magnetic field. In

Fig.15, the variation of the heat capacity (C) as a function of inverse temperature is shown. Fig.15 shows that the specific heat peaked at unity and decreased with increasing $\beta$ for different deformation parameter $q$ in the presence of the external magnetic field.

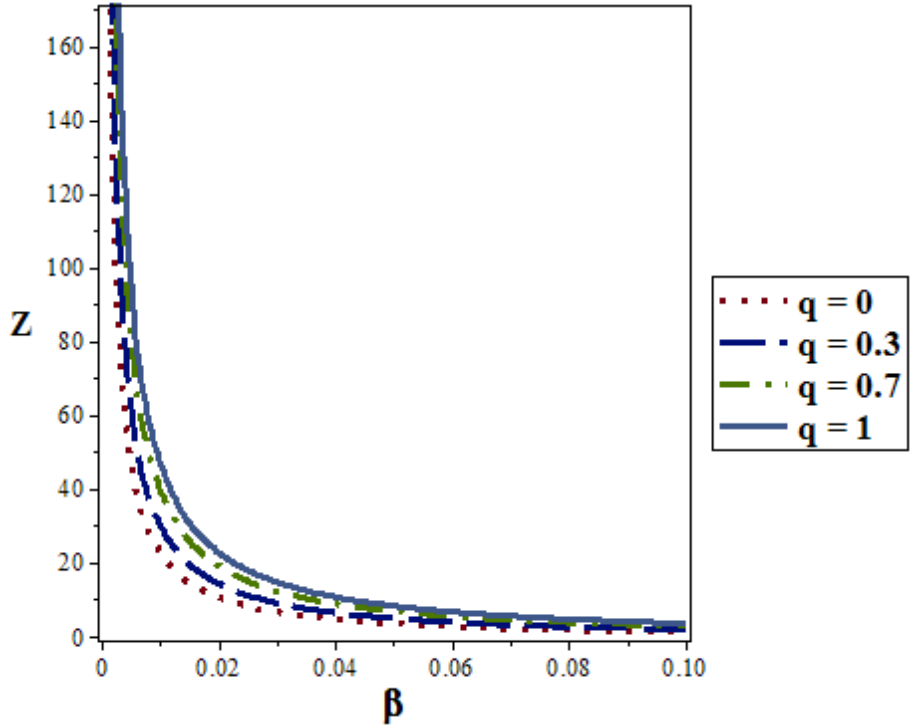

Fig. 11: Partition function vs $\beta$ for different deformation parameters.

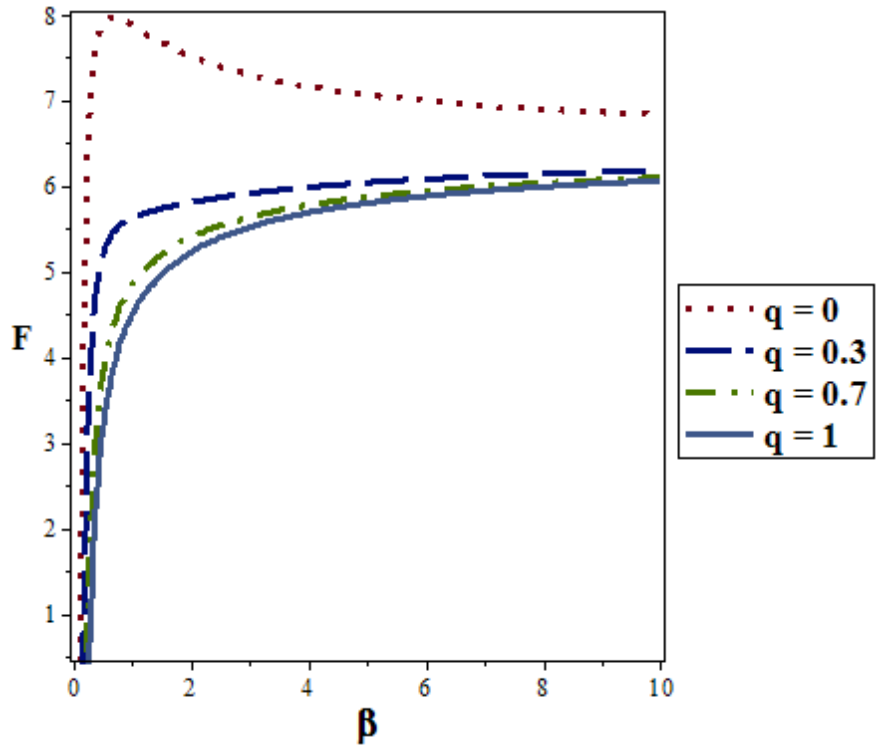

Fig:12: Helmholtz free energy vs $\beta$ for different deformation parameters

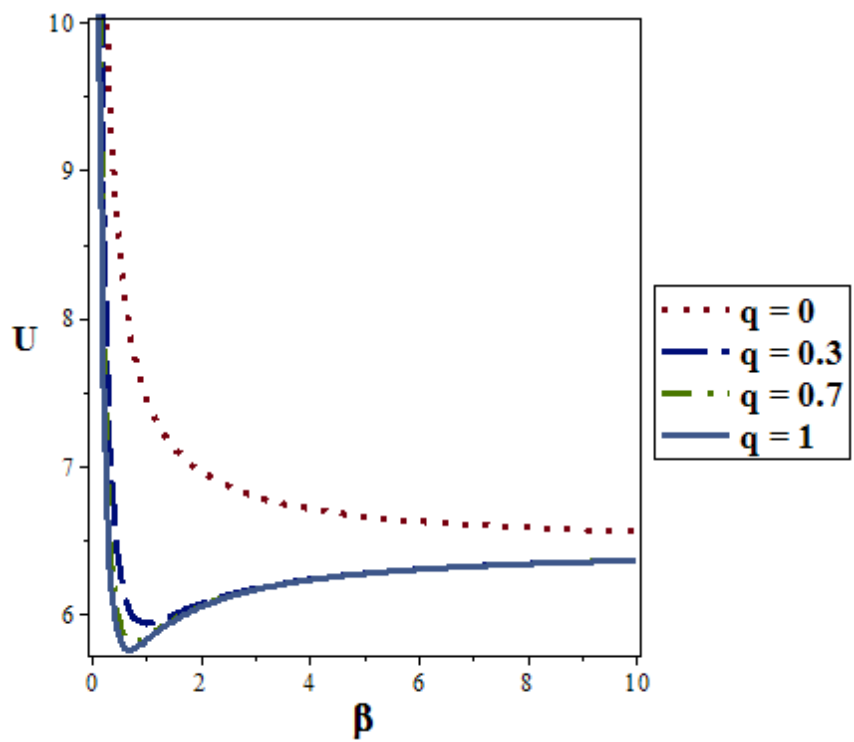

Fig. 13: Mean energy vs $\beta$ for different deformation parameters.

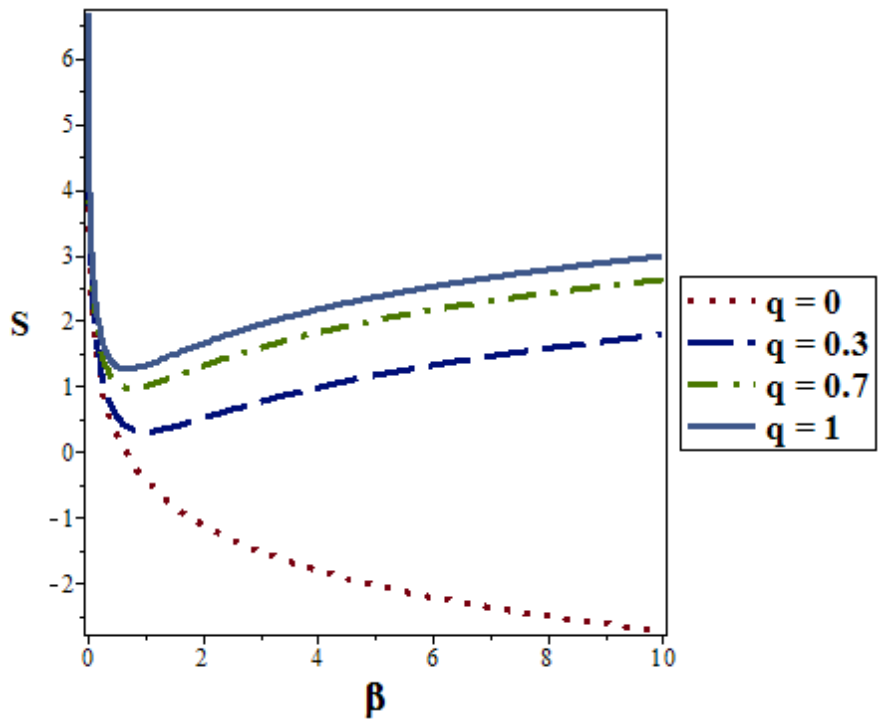

Fig. 14: Entropy vs $\beta$ for different deformation parameters.

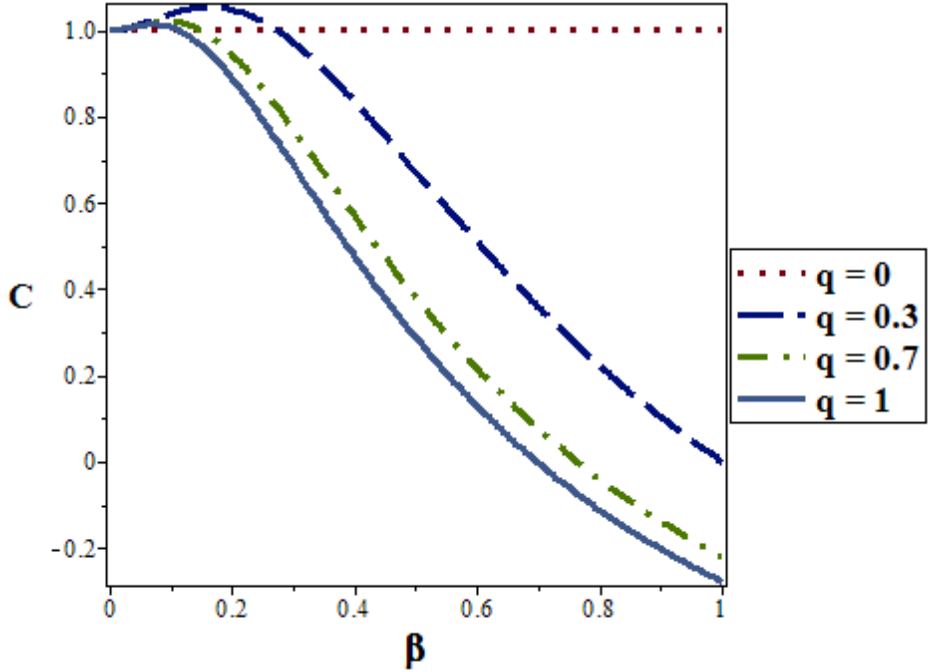

Fig. 15: Specific heat capacity vs $\beta$ for different deformation parameters.

## 5. Conclusions

In this work, we have studied the thermodynamic properties of pseudoharmonic potential in the presence of external magnetic and AB fields. We obtained in a closed form the energy spectrum of the system and calculate the partition function of the system within the framework of superstatistics using the modified Dirac delta distribution. With the partition function, we determine other thermodynamic properties such as Helmholtz free energy, mean energy, entropy and the specific heat capacity. We applied the result to study the thermodynamic properties of some selected diatomic molecules of $N_2, Cl_2, I_2$ and $CH$ within the framework of superstatistics .

## Appendix A: The parametric NU Method

The parametric form of the NU method takes the form [56]

$$\frac{d^2\psi}{ds^2} + \frac{\alpha_1 - \alpha_2 s}{s(1-\alpha_3 s)}\frac{d\psi}{ds} + \frac{1}{s^2(1-\alpha_3 s)^2}\left\{-\xi_1 s^2 + \xi_2 s - \xi_3\right\}\psi(s) = 0 \qquad (A.1)$$

The energy eigenvalues equation and eigenfunctions respectively satisfy the following sets of equation,

$$\alpha_2 n - (2n+1)\alpha_5 + (2n+1)(\sqrt{\alpha_9} + \alpha_3\sqrt{\alpha_8}) + n(n-1)\alpha_3 + \alpha_7 + 2\alpha_3\alpha_8 + 2\sqrt{\alpha_8\alpha_9} = 0,$$
(A.2)

$$\psi(s) = s^{\alpha_{12}}(1-\alpha_3 s)^{-\alpha_{12}-\frac{\alpha_{13}}{\alpha_3}} P_n^{(\alpha_{10}-1,\frac{\alpha_{11}}{\alpha_3}-\alpha_{10}-1)}(1-2\alpha_3 s) \qquad (A.3)$$

where

$$\alpha_4 = \frac{1}{2}(1-\alpha_1), \alpha_5 = \frac{1}{2}(\alpha_2 - 2\alpha_3), \alpha_6 = \alpha_5^2 + \xi_1,$$

$$\alpha_7 = 2\alpha_4\alpha_5 - \xi_2, \alpha_8 = \alpha_4^2 + \xi_3, \alpha_9 = \alpha_3\alpha_7 + \alpha_3^2\alpha_8 + \alpha_6$$

$$\alpha_{10} = \alpha_1 + 2\alpha_4 + 2\sqrt{\alpha_8}, \alpha_{11} = \alpha_2 - 2\alpha_5 + 2(\sqrt{\alpha_9} + \alpha_3\sqrt{\alpha_8})$$

$$\alpha_{12} = \alpha_4 + \sqrt{\alpha_8}, \alpha_{13} = \alpha_5 - (\sqrt{\alpha_9} + \alpha_3\sqrt{\alpha_8}) \qquad (A.4)$$

and $P_n$ is the orthogonal Jacobi-polynomial which define as

$$P_n^{(\alpha,\beta)}(x) = \frac{\Gamma(\alpha+n+1)}{n!\Gamma(\alpha+\beta+n+1)}\sum_{m=0}^{n}\binom{n}{m}\frac{\Gamma(\alpha+\beta+n+m+1)}{\Gamma(\alpha+m+1)}(\frac{x-1}{2})^m \qquad (A.5)$$

# References

[1] S.A.Najafizade, H.Hassanabdi and S.Zarrikamar, Chin.Phys.B 25(2016)040301
[2] S.Liu, J.Chem.Phys.126(2007)191107
[3] G.H.Sun, S.H.Dong and N.Saad, Ann.Phys.(Berlin) doi:10.1002/andp.201300089
[4] R.V.Torres, G.H.Sun and S.H.Dong, Phys.Scr.90(2015)035205
[5] G.H.Sun, M.Avila Aoki and S.H.Dong, Chin.Phys.B 22(2013) 050302
[6] A.N.Ikot E.O. Chukwuocha, M.C Onyeaju, C.A Onate, B.I. Ita and M.E.Udoh, Pramana-J Phys. 90(2018)22
[7] U.S.Okorie, E.E. Ibekwe, A.N.Ikot, M.C.Onyeaju and E.O.Chukwuocha, J Kor Phys Soc 79(2018)1211
[8] U.S.Okorie, A.N.Ikot, M.C.Onyeaju and E.O.Chukwuocha EO (2018) Rev Mex De Fis 64(2018)608
[9] U.S. Okorie, A.N. Ikot, M.C. Onyeaju and E.O.Chukwuocha EO (2018) J.Mol Mod 24(2018)289
[10] C.S.Jia, L.-H.Zhang C.W.Wang, Chem. Phys. Letts.667 (2017) 211
[11] S. Sargolzaeipor, H.Hassanabadi H and W.S Chung, Eur Phys J Plus 133(2018)5
[12] H.Sobhani, H.Hassanabadi and W.S. Chung, Eur Phys J C 78(2018)106
[13] S.Sargolzacipor, H.Hassanabadi and W.S.Chung,, Physica A, doi:10.1016/j.physa.2018.05.125(2018)
[14] S.Sargolzacipor, H.Hassanabadi and W.S.Chung, Mod.Phys.Lett.33 (2018)1850060
[15] A. N. Ikot, U.S.Okorie, C.A.Onate, M.C.Onyeaju and H. Hassanabad, Can.J.Phys. https://doi.org/10.1139/cjp-2018-0535
[16] P. Gupta and I. Mechrotra, J . Mod. Phys. 3(2012) 1530 .
[17] A. Al-Jamel and H. Widyan, Appl. Phys. Rese. 4(2013)94.
[18] A. N. Ikot, B. C. Lutfuoglu, M. I. Ngwueke, M. E. Udoh, S. Zare, and H. Hassanabadi, Eur. Phys. J. Plus, 131(2016) 419
[19] W.Greiner, Relativistic Quantum Mechanics(Berlin,Springer,2000)
[20] S.H.Dong, Wave Equations in Higher Dimensions(Springer-Verlag,New York,2011)
[21] S.M.Ikhdair and R.Sever, Theochem 855(2008)13
[22] S.M.Ikdair, B.J.Falaye and M.Hazamvi, Ann.Phys.353(2015)282

[23] S.M.Ikhdair and R.Sever,Theochem 806(2007)155

[24] H.Hassanabadi,E.Maghsoodi and S.Zarrinkamar,Ann.Phys(Berlin) (2013),doi:10.1002/andp.2013001202

[25] S.M.Ikhadiar and B.J.Falaye,J.Ass.Arab Univ.Bas.App.Sci. (2013),doi:10.1006/j.jaubas.2013.07.004

[26] M.Eshshi and H.Mehraban,Eur.Phys.J.Plus 132(2017)121

[27] M.Eshshi, H.Mehraban and S.M.Ikhdair,Eur.Phys.J.A 52(2016)201

[28] S.M.Ikhdair and R.Sever,Adv.High Ener.Phys. http://dx.doi.org/10.1155/2013/562959

[29] P.Koscik and A.Okopinska,J.Phys.A:Math.Theor.40(2007)1045

[30] M.Aygun,O.Bayrak,I.Boztosun and Y.Sahin,Eur.Phys.J.D 66(2012)35

[31] H. Ciftci, R. L. Hall, N. Saad, *J. Phys. A: Math Gen*. **36** (2003)11807

[32] S.M.Ikhdair and M.Hamzavi,Physica B 407(2012)4198

[33] A. F. Nikiforov and V. B. Uvarov, Special functions of Mathematical Physics (BirkhauserVerlag, Basel,1988).

[34] A.A.Rajabi and M.Hamzavi,Eur.Phys.J.Plus 128(2013)5

[35] H.Hamzavi and S.M.Ikhdair,Mod.Phys.Letts.B 27(2013)1350176

[36] M.Eshghi and S.M.Ikhdair,Chin.Phys.B 27(2018)080303

[37] S.M.ikhdair and M.Hamzavi,Chin.Phys.B 21(2012)110302

[38] R.Khordad,M.A.Sadeghzadeh and A.M.Jahan-abad, Commun.Theor.Phys.59(2013)655

[39] J.D.Castano-Yepes,C.F.R.Gutierrez,H.C.Gallego and E.A.Gomez, Physica E,103(2018)464

[40] J.I.Climente,J.Planelles and J.L.Movilla,Rev.B 46(1992)12773

[41] S.M.Reimann and M.Manninen,Rev.Mod.Phys.74(2002)1283

[42] M.S.Atoyan,E.M.Kazaryan and H.A.Sarkisyan,Physica E 31(2014)1

[43] Sukirti Gumber, Manoj Kumar, Monica Gambhir, Man Mohan, and Pradip Kumar Jha, Can.J.Phys.93(2015)1

[44] W.Ebeling and M.I.Sokolov,Statistical Thermodynamics and Stochastic Theory of Nonequilibrium Systems(World Scientific,Singapore,2005).

[45] J.W.Gibbs,Collected Works and Commentary Vol.1 and II,Yale University Press (1936)

[46] A.Einstein,Annalen der Physikb14(1904)354

[47] P.T.Landsberg,Thermodynamics and Statistical Mechanics (Dover,New York 1991)

[48] C.Tsallis,J.Stat.Phys.52(1998)479

[49] G.Wilk and Z.Wlodarczyk,Phys.Rev.Lett.84(2002)2770

[50] C.Beck and E.G.D.Cohen,Physica A 322(2003)267

[51] C.Beck Physica A 331(2004)173

[52] S.Rizzo and A.Rapisarda,in Experimental Chaos:8[th] Experimental Chaos Conference, edited by S.Boccaletti,D.Yardanov, R.Meucci, L.M.Pecora,J.Kurths and B.J.Gluckman,AIP Conf.Proc.No.742 (AIP,Melville,NY ),176(2004).
[53] A.M.Reynolds,Phys.Rev.Lett.91(2003)084503
[54] A.Cetin,Phys.Lett.A 372(2008)3852
[55] C.Birkdemir,A.Birkdemir and J.Han,Chem.Phys.Lett.417(2006)326
[56] C.Tezcan and R.Sever, Int J Theor Phys 48(2009)337